\documentstyle{article}

\def\be{\begin{equation}}
\def\ee{\end{equation}}
\def\bea{\begin{eqnarray}}
\def\eea{\end{eqnarray}}
\def\Ref#1{(\ref{#1})}

\begin{document}
\begin{titlepage}
\vspace*{10mm}
\begin{center}
{\large \bf Annihilation--diffusion processes: an exactly solvable model}
\vskip 10mm
\centerline {\bf Farinaz Roshani $^a$ {\rm and} Mohammad Khorrami $^b$}
\vskip 1cm
{\it Institute for Advanced Studies in Basic Sciences,}
\\ {\it P. O. Box 159, Zanjan 45195, Iran}\\
{\it Institute for Studies in Theoretical physics and Mathematics,}\\
{\it P. O. Box 5531, Tehran 19395, Iran}\\
$^a$ roshani@iasbs.ac.ir\\
$^b$ mamwad@iasbs.ac.ir\\

\end{center}
\vskip 2cm
\begin{abstract}
\noindent A family of diffusion--annihilation processes is introduced,
which is exactly solvable. This family contains parameters that control the
diffusion- and annihilation- rates. The solution is based on the Bethe
ansatz and using special boundary conditions to represent the reaction. 
The processes are investigated, both on the lattice and on the continuum.
Special cases of this family of processes are the simple exclusion process
and the drop--push model.
\end{abstract}
\end{titlepage}
\newpage
\section{Introduction}
In recent years, the asymmetric exclusion process and the problems related
to it, including for example the bipolimerization \cite{1}, dynamical models
of interface growth \cite{2}, traffic models \cite{3}, the noisy Burgers
equation \cite{4}, and the study of shocks \cite{5,6}, have been extensively
studied. The dynamical properties of this model have been studied in [6--8].
As the results obtained by approaches like mean field are not reliable in
one dimension, it is useful to introduce solvable models and analytic
methods to extract exact physical results. Among these methodes is the
coordinate Bethe--ansatz, which was used in \cite{Sc} to solve the
asymmetric simple exclusion process on a one--dimensional lattice. In
\cite{AKK1}, a similar technique was used to solve the drop--push model
\cite{11}, and a generalized one--parameter model interpolating between the
asymmetric simple exclusion model and the drop--push model. In \cite{AKK2},
this family was further generalized to a family of processes with arbitrary
left- and right- diffusion rates. All of these models were lattice models.
Finally, the behaviour of latter model on continuum was investigated in
\cite{RK}. The continuum models of this kind are also investigated in
\cite{SW1,SW2}.

In the generalized model interpolating between the asymmetric simple
exclusion model and the drop--push model \cite{AKK1,AKK2,RK}, there are
two paprameters $\lambda$ and $\mu$, which control the pushing rate.
Normalizing the diffusion rate to one, it is seen that the sum of these two
parameters should be one to ensure the conseravtion of probability. These
two parameters appear in the boundary condition used instead of the
reaction. A question arises that what happens if one violates this
conservation of probability. This is what is investigated in the present
paper.

The scheme of the paper is the following. In section 2, the allowed boundary
conditions are investigated. It is shown that if $\lambda +\mu <1$, then the
number of the particles will be decreasing, that is, there is an
annihilation process as well. It is shown that one can in fact write a
two--particle to one--particle annihilation process which results in such a
boundary condition.

In section 3, the Bethe--ansatz solution for the $N$--particle probability
of this process is obtained, and its large--time behaviour is investigated.
This is done for the process on the lattice as well as on the continuum.
Finally, in section 4 the special case of the two--particle initial
condition is fully investigated, and it is explicitely shown that at
large times, there remains only one particle.

\section{Boundary conditions}
Consider the following master equation for an asymmetric exclusion process.
\bea\label{1}
{\partial\over{\partial t}}P(x_1,x_2,\cdots ,x_N;t)&=&
P(x_1-1,x_2,\cdots ,x_N;t)\cr
&&+P(x_1,x_2-1,\cdots ,x_N;t)+\cdots\cr
&&+P(x_1,x_2,\cdots ,x_N-1;t)\cr
&&-NP(x_1,x_2,\cdots ,x_N;t).
\eea
This equation describes a collection of $N$ particles drifting to the
right with unit rate. If the particles are to exclude each other, that is
if no two particles are to occupy the same site, then \Ref{1} is valid only
for
\be
x_i<x_{i+1}-1.
\ee
One can, however, assume that \Ref{1} is correct for all of the physical
region $x_i<x_{i+1}$, and impose certain boundary conditions for
$x_i=x_{i+1}$. Note that if $x_i=x_{i+1}-1$ for some $i$, then in the
right--hand side of \Ref{1} there will be terms with $x_i=x_{i+1}$, which
is out of the physical region. The boundary condition determines the
nature of the interaction between particles. But what are the allowed
boundary conditions? Let's rewrite \Ref{1} for the case of two particles
and use the conservation of probability. We arrive at
\bea\label{3}
{\partial\over{\partial t}}\sum_{x_2}\sum_{x_1<x_2}P(x_1,x_2;t)&=&
\sum_{x_2}\sum_{x_1<x_2}P(x_1,x_2;t)\cr
&&-\sum_xP(x,x+1;t)\cr
&&+\sum_{x_2}\sum_{x_1<x_2}P(x_1,x_2;t)\cr
&&+\sum_xP(x,x;t)\cr
&&-2\sum_{x_2}\sum_{x_1<x_2}P(x_1,x_2;t),\cr
&=&-\sum_xP(x,x+1;t)+\sum_xP(x,x;t).
\eea
If the right--hand side of \Ref{3} is to be identical to zero, then 
$P(x,x)$ should be a linear combination of $P(x-1+i,x+i)$'s. This may work
for the two--particle process, but in many--partcles processes it may
introduce terms with $x_i>x_{i+1}$, which need additional boundary
conditions. The only exception is when $P(x,x)$ is a linear combination of
only $P(x,x+1)$ and $P(x-1,x)$. So one can write
\be\label{4}
P(x,x)=\lambda P(x,x+1)+\mu P(x-1,x).
\ee
Inserting this in \Ref{3}, one arrives at 
\be\label{5}
{\partial\over{\partial t}}\sum_{x_2}\sum_{x_1<x_2}P(x_1,x_2;t)=
(\lambda +\mu -1)\sum_xP(x,x+1;t).
\ee
In order that the right--hand side of \Ref{5} be zero, one must impose
\be\label{6}
\lambda +\mu =1.
\ee
This is the boundary condition used in \cite{AKK1,AKK2,RK}.

This kind of boundary condition ensures the conservation of particle number.
But in a process where annihilation exists as well, the number of the
particles is not conserved; it is decreasing. It is seen that if
\be\label{7}
\lambda +\mu <1,
\ee
then the probability of finding two particles is decreasing. Suppose one
begins with two particles. They drift to right with unit rate. If they meet
each other, either the left particle is stopped, or one of them is
annihilated. That is, we have the following processes:
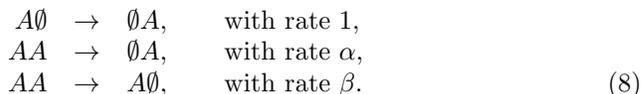
\bea\label{7*}
A\emptyset&\to&\emptyset A,\qquad\hbox{with rate }1,\cr
AA&\to&\emptyset A,\qquad\hbox{with rate }\alpha ,\cr
AA&\to&A\emptyset ,\qquad\hbox{with rate }\beta .
\eea
In this case, still the more--than--two particle densitiies are zero, since
no particles are generated during the process. But the two--particle density
does not determine the one--particle density. And the summation of the
former need not be a constant (one). 
It is seen that the master equation for the two--particle probability is
\bea\label{8}
{\partial\over{\partial t}}P(x_1,x_2;t)&=&P(x_1-1,x_2;t)+P(x_1,x_2-1;t)\cr
&&-2 P(x_1,x_2;t),\qquad\qquad x_1<x_2-1,
\eea
and
\be\label{9}
{\partial\over{\partial t}}P(x,x+1;t)=P(x-1,x+1;t)-(1+\alpha +\beta )
P(x,x+1;t).
\ee
But \Ref{9} is the same as \Ref{8}, provided one uses the boundary condition
\be\label{11}
P(x,x)=\lambda P(x,x+1),
\ee
with
\be
\lambda =1-(\alpha +\beta ).
\ee
So the difference $1-\lambda$ is in fact related to the annihilation rate,
as expected. There is one other thing to be noted. As the number of
particles is not conserved, one cannot calculate the one--particle
probability by a simple summation of the two--particle probability. That is,
\be
P(x)\ne\sum_{y>x}P(x,y)+\sum_{y<x}P(y,x).
\ee
In fact, for the process described, the particles interact and annihilate
each other, until there remains only one particle. This means that at
$t\to\infty$, there will be only one particle. So the
more--than--one--particle probabilities will tend to zero, whereas the
summation of the one--particle probability tends to one. However, if the
initial condition is that there are $N$ particles, one can write
differential equations for $n$--particle probabilities in which
$n$--particle probabilities and $n+1$--particle probabilities occur (if
$n<N$). For $n>N$, the $n$--particle probability is identically zero, and
the equation for the $N$--particle probability is closed. So, in principle,
one can find the $N$--particle probability first and use this to find
less--than--$N$--particle probabilities. To be specific, the evolution
equation for the one--particle probability is
\bea\label{14}
{\partial\over{\partial t}}P(x;t)&=&P(x-1;t)[1-P(x;t)]-P(x;t)[1-P(x+1;t)]
\cr
&&-\alpha P(x,x+1;t)-\beta P(x-1,x;t).
\eea

\section{Bethe--ansatz solution for the $N$--particle probability}
Consider the Master equation \Ref{1}, with the boundary condition
\be
P(\cdots ,x,x,\cdots )=\lambda P(\cdots ,x,x+1,\cdots ),
\ee
where $\lambda <1$. Following \cite{Sc,AKK1,AKK2,RK}, one can obtain the
conditional probability using the Bethe ansatz. Writing
\be\label{16}
P({\bf x};t)=e^{Et}\Psi ({\bf x}),
\ee
and
\be\label{17}
\Psi ({\bf x})=\sum_\sigma A_\sigma e^{i\sigma ({\bf p})\cdot{\bf x}},
\ee
where the summation runs over the elements of the permutation group, one
arrives at
\be\label{18}
E=-N+\sum_j e^{-ip_j},
\ee
and
\be
A_{\sigma\sigma_i}=S[\sigma (p_i),\sigma (p_{i+1})]A_\sigma,
\ee
where $\sigma$ is that permutation which only interchanges $p_i$ and
$p_{i+1}$. One also finds that
\be
S_{jk}:=S(p_j,p_k)=-{{1-\lambda e^{ip_k}}\over{1-\lambda e^{ip_j}}}.
\ee
This is the same as what found in \cite{AKK1,AKK2} with $\mu =0$, and if one
puts $\lambda =1$, the result of \cite{Sc} is obtained. The conditional
probability is thus written as
\be
P({\bf x};t|{\bf y};0)=\int{{d^N p}\over{(2\pi)^N}}\Psi_{\bf p}({\bf x})
e^{E({\bf p})t-i{\bf p}\cdot{\bf y}},
\ee
where $\Psi$ is defined as \Ref{17} with $A_{\rm identity}=1$. This looks
like similar to what obtained in \cite{Sc,AKK1,AKK2}. There is, however,
a difference. As $\lambda <1$, there is no pole in $S$, and hence in $A$.
So for large times, when the probability distribution becomes smooth and
its Fourier--transform for large frequencies tends to vanish, one can put
$p_j=0$ in $S$ as an approximation to arrive at
\be
S\approx -1,
\ee
and
\be
A_\sigma\approx (-1)^\sigma .
\ee
One can also approximate $E({\bf x})$ as
\be
E({\bf p})\approx\sum_j\left(-ip_j-{{p_j^2}\over 2}\right) .
\ee
So, for large times,
\be\label{20}
P({\bf x};t|{\bf y};0)\approx{1\over{(2\pi t)^{N/2}}}\sum_\sigma (-1)^\sigma
e^{-\sum_j[x_j-\sigma (y_j)-t]^2/(2t)}.
\ee
It is clearly seen that the integral of this distribution over the physical
region tends to zero as $t\to\infty$. This should be the case, since the
number of the particles does not remain constant and decreases.

Using the boundary condition \Ref{4} with \Ref{7}, doesn't change the
results drastically. In fact, $E$ doesn't change at all, while $S$ is
changed to
\be
S_{jk}=-{{1-\lambda e^{ip_k}-\mu e^{-ip_j}}\over{1-\lambda e^{ip_j}
-\mu e^{-ip_k}}}.
\ee
The approximate result for large times does not depend on $\lambda$ or
$\mu$, so long as their sum is less than 1.

One can also investigate the continuous--space form of the evolution.
Following \cite{RK}, the master equation is changed to
\be\label{25}
{\partial\over{\partial t}}P({\bf x};t)=-\sum_j\partial_j P({\bf x};t)
+{1\over 2}\sum_j\partial_j^2 P({\bf x};t),
\ee
and the boundary condition to
\be\label{26}
(1-\lambda -\mu-\lambda\partial_{j+1}+\mu\partial_j)P|_{x_{j+1}=x_j}=0.
\ee
Using the Galilean transformation $x_i\to x_i+vt$ and $t\to t$, the master
equation \Ref{25} is simplified to
\be\label{27}
{\partial\over{\partial t}}P({\bf x};t)={1\over 2}\nabla^2 P({\bf x};t),
\ee
Using a Bethe--ansatz solution like \Ref{16} and \Ref{17}, one arrives at
\be\label{30}
E=-{1\over 2}\sum_j p_j^2,
\ee
and
\be\label{31}
S_{jk}=-{{1-\lambda -\mu-i\lambda p_k+i\mu p_j}\over
         {1-\lambda -\mu-i\lambda p_j+i\mu p_k}}.
\ee
For large times, one can approximate $S$ to $-1$, and arrive at a result
similar to \Ref{20}. The difference is that in the exponent the term
$x_j-t-\sigma (y_j)$ is replaced by $x_j-\sigma (y_j)$, as the Galilean
transformation used has canceled the drift from the master equation.

\section{Two--particle system and the exact solution}
As it was seen in the previous section, the conditional probability for the
two--particle system described by \Ref{8} and \Ref{11} is
\bea\label{32}
P({\bf x};t|{\bf y};0)&=&\int{{d^2p}\over{4\pi^2}}e^{Et-i{\bf p}
\cdot{\bf y}}\cr
&&\times\left[e^{i(p_1x_1+p_2x_2)}-{{1-\lambda e^{ip_2}}\over
{1-\lambda e^{ip_1}}}e^{i(p_1x_2+p_2x_1)}\right] ,
\eea
where $E$ is obtained from \Ref{18}. This integration is easily done and the
result is
\bea\label{33}
P({\bf x};t|{\bf y};0)&=&e^{-2t}{{t^{x_1-y_1}}\over{(x_1-y_1)!}}
{{t^{x_2-y_2}}\over{(x_2-y_2)!}}\cr
&&+e^{-2t}\sum_{l=0}^\infty{{t^{l+x_2-y_1}}\over{(l+x_2-y_1)!}}
{{t^{x_1-y_2}}\over{(x_1-y_2)!}}\lambda^l
\left(-1+{{\lambda t}\over{x_1-y_2+1}}\right) .
\eea

Another interesting quantity is the average number of the particles. This
is equal to the summation of the one--particle probability:
\be\label{34}
N(t):=\sum_x P(x;t).
\ee
Using \Ref{14}, one arrives at
\bea\label{35}
\dot N&=&-(\alpha +\beta)\sum_x P(x-1,x;t)\cr
      &=&-{{1-\lambda}\over\lambda}\sum_xP(x,x;t).
\eea
The right--hand side can be calculated using the Bethe--ansatz solution
directly. Using \Ref{32}, one has
\bea\label{36}
\sum_x P(x,x;t)&=&\int{{d^2p}\over{2\pi}}\delta (p_1+p_2)[1+S(p_1,p_2)]
e^{Et-i(p_1y_1+p_2y_2)}\cr
&=&\lambda\int{{dp}\over{2\pi}}e^{2t(\cos p-1)+ip(y_2-y_1)}{{e^{-ip}-e^{ip}}
\over{1-\lambda e^{ip}}}\cr
&=&\lambda\int{{dp}\over{2\pi}}e^{2t(\cos p-1)+ip(y_2-y_1)}\cr
&&\times\sum_{m=0}^\infty\lambda^m\left[ e^{i(m-1)p}-e^{i(m+1)p}\right]\cr
&=&\lambda\sum_{m=0}^\infty e^{-2t}\lambda^m\left[{\rm I}_{y_2-y_1+m-1}(2t)
-{\rm I}_{y_2-y_1+m+1}(2t)\right]\cr
&=&\lambda\sum_{m=0}^\infty e^{-2t}\lambda^m{{y_2-y_1+m}\over t}
{\rm I}_{y_2-y_1+m}(2t),
\eea
where I denotes the modified Bessel function. This can be inserted in
\Ref{35} to obtain
\bea\label{37}
N(t)&=&N(0)-{{1-\lambda}\over\lambda}\int_0^tdt'\sum_x P(x,x;t')\cr
&=&N(0)-(1-\lambda)\sum_{m=0}^\infty\int_0^t dt' e^{-2t'}\lambda^m
{{y_2-y_1+m}\over t'}{\rm I}_{y_2-y_1+m}(2t').
\eea
This is simplified for $t\to\infty$. Using
\be\label{38}
\int_0^\infty ds{{e^{-s}}\over s}{\rm I}_n(s)={1\over n},
\ee
one arrives at
\bea\label{39}
N(\infty)&=&N(0)-(1-\lambda )\sum_{m=0}^\infty\lambda^m,\cr
&=&N(0)-1.
\eea
But note that
\be\label{40}
N(0)=2,
\ee
since at the beginning there were two particles at $y_1$ and $y_2$. That is,
\be\label{41}
P(x;0)=\delta (x-y_1)+\delta (x-y_2).
\ee
So, at $t\to\infty$, there remains only one particle, as one of the two
particles has been annihilated.

The continuous--space analogue of this model can also be solved easily.
Using \Ref{27} as the master equation, and \Ref{30} and \Ref{31} (with
$\mu =0$), one is led to
\bea\label{42}
P({\bf x};t|{\bf y};0)&=&\int{{d^2p}\over{4\pi^2}}e^{Et-i{\bf p}
\cdot{\bf y}}\cr
&&\times\left[e^{i(p_1x_1+p_2x_2)}-{{1-\lambda-i\lambda p_2}\over
{1-\lambda-i\lambda p_1}}e^{i(p_1 x_2+p_2 x_1)}\right].
\eea
Using the change of variable $p:=p_1+i(1-\lambda )/\lambda$ in the second
integral, \Ref{42} is written as
\bea\label{43}
P({\bf x};t|{\bf y};0)&=&{1\over{2\pi t}}e^{-[(x_1-y_1)^2+(x_2-y_2)^2]/(2t)}
\cr
&&+{1\over{\sqrt{8\pi t}}}\left[ A+{{x_1-y_2}\over t}\right]
e^{-(x_1-y_2)^2/(2t)}e^{[2A(x_2-y_1)+tA^2]/2}\cr
&&\times\left\{ -1+{\rm erf}\left[{1\over{\sqrt{2t}}}(x_2-y_1+tA)\right]
\right\} ,
\eea
where
\be\label{44}
A:={{1-\lambda}\over\lambda}.
\ee
One notes that at $t\to\infty$, the conditional probability is simplified.
We have
\be\label{45}
1-{\rm erf}(x)\approx{{e^{-x^2}}\over{x\sqrt{\pi}}},\qquad\hbox{for large
$x$}
\ee
From this, it is seen that at $t\to\infty$
\be\label{46}
P({\bf x};t|{\bf y};0)\approx {1\over{2\pi t}}\left\{ e^{-[(x_1-y_1)^2+
(x_2-y_2)^2]/(2t)}-e^{-[(x_1-y_2)^2+(x_2-y_1)^2]/(2t)}\right\} .
\ee
This is a special case of what obtained in the previous section.

Another quantity to be considered is the one--point probability. In the
continuum limit, and after performing the Galilean transformation, \Ref{14}
becomes
\be\label{47}
{{\partial}\over{\partial t}}P(x;t)={1\over 2}{{\partial^2}\over
{\partial x^2}}P(x;t)-[\alpha (1+\partial_2)+\beta (1-\partial_1)]P(x,x;t).
\ee
From this, using the boundary condition \Ref{26}, with $\mu =0$, one arrives
at
\be\label{48}
\dot N(t)=-{{1-\lambda}\over\lambda}\int dx\; P(x,x;t),
\ee
where
\be\label{49}
N(t):=\int dx\; P(x;t).
\ee
Using \Ref{42}, the integral at the right--hand side is calculated to be
\be\label{50}
\int dx\; P(x,x;t)=\int{{dp}\over{2\pi}}{{-2i\lambda p}\over
{1-\lambda -i\lambda p}}e^{-tp^2+ip(y_2-y_1)}.
\ee
To obtain $N(\infty)$, one integrates \Ref{50} from $0$ to $\infty$. This
results in
\bea\label{51}
\int_0^\infty dt\int dx\; P(x,x;t)&=&{\lambda\over\pi}{\rm P}\int{{dp}\over
{ip}}{{e^{ip(y_2-y_1)}}\over{1-\lambda -i\lambda p}}\cr
&=&{\lambda\over{1-\lambda}}.
\eea
The symbol P denotes the Cauchy's principle value, and use has been made of
the fact that $y_2>y_1$. From this, it is found that
\bea\label{52}
N(\infty)&=&N(0)-1\cr
&=&1.
\eea
This is the same result obtained for the lattice, as expected.

\end{document}